\RequirePackage[mathlines]{lineno} % Display line numbers
\documentclass[prd,twocolumn,showpacs,amsmath,amssymb,aps]{revtex4-1}
\usepackage[colorlinks,linkcolor=blue,anchorcolor=blue,citecolor=blue,pdfborder={0 0 0}]{hyperref}
\usepackage{overpic,graphicx}% Include figure files
\usepackage{dcolumn}% Align table columns on decimal point
\usepackage{bm}% bold math
\usepackage{rotating}
\usepackage{times}
\usepackage{amsfonts}
\usepackage{amsthm}
\usepackage{latexsym}
\usepackage{color}
\usepackage{url}
\usepackage{subfigure}
\usepackage{float}
\usepackage{mathrsfs}
\usepackage{multirow}
\begin{document}
\normalsize
%%\linenumbers
\hyphenpenalty=5000
\tolerance=1000
\lefthyphenmin=2
\righthyphenmin=2
\uchyph=0
\frenchspacing

\newcommand{\jpsi}{J/\psi}
\newcommand{\etap}{\eta^{\prime}}
\newcommand{\ub}{\gamma^{\prime}}
\newcommand{\picw}{0.36\textwidth}
\newcommand{\cred}{\textcolor{red}}
\newcommand{\mee}{m_{e^+e^-}}

\title{ \boldmath Measurement of $\mathcal{B}(J/\psi \to \etap e^+ e^- $) and search for a  dark photon }

\author{
\begin{small}
\begin{center}
  M.~Ablikim$^{1}$, M.~N.~Achasov$^{9,d}$, S. ~Ahmed$^{14}$,
  M.~Albrecht$^{4}$, M.~Alekseev$^{55A,55C}$, A.~Amoroso$^{55A,55C}$,
  F.~F.~An$^{1}$, Q.~An$^{52,42}$, J.~Z.~Bai$^{1}$, Y.~Bai$^{41}$
% ...
  \\
  \vspace{0.2cm}
  (BESIII Collaboration)\\
  \vspace{0.2cm} {\it
    $^{1}$ Institute of High Energy Physics, Beijing 100049, People's Republic of China\\
% ...
$^{4}$ Bochum Ruhr-University, D-44780 Bochum, Germany\\
% ...
$^{9}$ G.I. Budker Institute of Nuclear Physics SB RAS (BINP), Novosibirsk 630090, Russia\\
% ...
$^{14}$ Helmholtz Institute Mainz, Johann-Joachim-Becher-Weg 45, D-55099 Mainz, Germany\\
% ...
$^{41}$ Southeast University, Nanjing 211100, People's Republic of China\\
    $^{42}$ State Key Laboratory of Particle Detection and Electronics, Beijing 100049, Hefei 230026, People's Republic of China\\
% ...
$^{52}$ University of Science and Technology of China, Hefei 230026, People's Republic of China\\
% ...
$^{55}$ (A)University of Turin, I-10125, Turin, Italy; (B)University of Eastern Piedmont, I-15121, Alessandria, Italy; (C)INFN, I-10125, Turin, Italy\\
% ...
$^{d}$ Also at the Novosibirsk State University, Novosibirsk, 630090, Russia\\
% ...
  }\end{center} \vspace{0.4cm}
\end{small}
}
%\date{\today}
\begin{abstract}
  Using a data sample of $(1310.6\pm7.0)\times10^{6}$ $J/\psi$ decay
  events collected with the BESIII detector at BEPCII, we study the
  electromagnetic Dalitz decay $\jpsi \to \etap e^+e^-$ with two
  dominant $\etap$ decay modes, $\etap \to \gamma \pi^+ \pi^-$ and
  $\etap \to \pi^+\pi^-\eta$. The branching fraction is determined to
  be
  $\mathcal{B}(\jpsi \to \etap e^+e^-) = (6.59\pm0.07\pm0.17) \times
  10^{-5}$, which improves in precision by a factor of 2 over the
  previous BESIII measurement. A search for the dark photon ($\ub$) is
  performed via $J/\psi \to\etap \ub, \ub \to e^{+}e^{-}$.  Excluding
  the $\omega$ and $\phi$ mass regions, no significant signal is
  observed in the mass range from 0.1 to 2.1 GeV/$c^{2}$. We set upper
  limits at the 90\% confidence level on
  $\mathcal{B}(\jpsi \to \etap \ub)\times\mathcal{B}(\ub \to e^+e^-)$,
  $\mathcal{B}(J/\psi \to\etap \ub$) and the mixing strength as a
  function of dark photon mass. This is among the first searches for dark
  photons in charmonium decays.
\end{abstract}

\maketitle

\section{\boldmath Introduction}

The electromagnetic (EM) Dalitz decay of a vector meson ($V$) to a
pseudoscalar meson ($P$) and a pair of leptons ($l= e , \mu$),
$V \to P l^+l^-$, provides important information on the interaction at
the $V$-$P$ transition vertex~\cite{r_lans}, where the lepton pair in
the final state originates from a virtual photon. Such Dalitz
processes have been widely studied with light unflavored meson decays,
such as $\phi \to \pi^0e^+e^-$~\cite{kloe_phipi0ee},
$\omega \to \pi^0 e^+e^-$~\cite{cmd2_omegapi0ee,a2_omegapi0ee},
$\omega \to \pi^0 \mu^+\mu^-$~\cite{na60_omegapi0mumu} and
$\phi \to \eta e^+e^-$~\cite{snd_phietaee,kloe_phietaee}. BESIII
observed the decays $\jpsi \to Pe^+e^-$~\cite{r_chuxkpee} for the
first time using $(225.3 \pm 2.8)\times 10^{6}$ $\jpsi$ events. The
branching fraction (BF) of $\jpsi \to \etap e^+e^-$ was measured to be
$(5.81\pm0.16\pm0.31)\times 10^{-5}$. It agrees with the theoretical
prediction $(5.66\pm0.16)\times 10^{-5}$~\cite{r_fjl} within the
uncertainty.
%The total
%sample of $(1310.6\pm7.0)\times10^{6}$ $J/\psi$ events at BESIII can
%be used to update the branching fraction of $\jpsi \to \etap e^+e^-$
%and to search for new phenomena.

Except for gravitational effects, we still know very little about the
constituents and interactions of dark matter. Many models beyond the
Standard Model (SM) of particle physics have proposed the existence of
a dark sector, which is being searched for with efforts from different
types of experiments~\cite{dp_babar1, dp_a1, dp_apex, dp_kloe, dp_a12,
dp_babar2, dp_na48, dp_kloe2, dp_kloe3, dp_bes3, dp_lhcb}.  The simple realizations of these
models usually consist of an extra U(1) gauge group, with a
corresponding massive vector boson force carrier, called a dark photon
($\ub$), which is neutral under the SM gauge symmetries, but couples
to the SM photon via kinetic mixing~\cite{r_1986} and decays into SM
particles. Such models provide a natural scenario for dark matter
interactions. A dark photon with a mass in the MeV/$c^2$ to GeV/$c^2$
range can also be accommodated by observational astroparticle
anomalies~\cite{r_nima}. Low-energy electron-positron colliders offer
an ideal environment to test these low-mass dark sector
models~\cite{r_Essig,Batell2009}, and meson decays provide an
important constraint on the mixing strength $\varepsilon$ between the
dark photon and SM photon~\cite{r_fayet07,wang_gev}. The authors of
Ref.~\cite{r_fjl} have estimated the achievable limits on the mixing
strength in the processes $\jpsi \to P\gamma' (\gamma'\to l^+l^-)$
using the huge BESIII $\jpsi$ data sample. The search in $\jpsi$
decays could uniquely probe the coupling of the dark photon with the charm
quark.

In this paper, we report on the updated BF measurement of
$\jpsi \to \etap e^+e^-$ and a search for a dark photon through
$\jpsi \to \etap \ub,\ub \to e^{+}e^{-}$, with 405 ${\rm pb}^{-1}$
$e^+ e^-$ collision data containing $(1310.6\pm7.0)\times10^{6}$
$J/\psi$ events~\cite{r_jpsi_no} collected by BESIII. Together with
the study of $\jpsi \to \eta e^+e^-$~\cite{r_etaee} with the same
data set, it is the first time that the dark photon is searched for
through the charmonium decays.

\section{\boldmath Apparatus and Monte Carlo simulation}
\label{bes3nmc}
BEPCII is a double ring
$e^+e^-$ collider running at the center-of-mass (c.m.) energy
$\sqrt{s}$ from 2.0 to 4.6 GeV with a peaking luminosity of
1$\times$10$^{33}$
cm$^{-2}$s$^{-1}$. The BESIII detector~\cite{r_bes3nim}, with a
geometrical acceptance of 93\% of the
4$\pi$ stereo angle, operates in a magnetic field of 1.0~T (0.9~T in
2012) provided by a superconducting solenoid. It is composed of a
helium-based main drift chamber (MDC) to measure the momentum and
ionization energy loss ($dE/dx$) of charged particles, a plastic scintillator time-of-flight
(TOF) system for particle identification (PID) information, a CsI(Tl)
electromagnetic calorimeter (EMC) to measure photon and electron
energies and a multilayer resistive plate chamber muon detection
system to identify muons.

Monte Carlo (MC) simulations are used to optimize the event selection,
investigate background and determine the detection efficiency. The
{\sc geant}4-based~\cite{r_geant4} simulation includes the description
of the geometry and material of the BESIII detector, the detector
response, and digitization models and also tracks the detector running
conditions and performance. An inclusive MC sample containing $1.225
\times 10^9$ $\jpsi$ events is used to study potential
backgrounds. The production of the $\jpsi$ meson is simulated by
the MC event generator {\sc kkmc}~\cite{r_kkmc}. The known decay modes
of the $\jpsi$ are generated by {\sc evtgen}~\cite{r_evtgen} with
BFs set at the world average values from the particle data
group (PDG)~\cite{r_pdg16}, while the remaining unknown decays are
generated by {\sc lundcharm}~\cite{r_lunccharm}. The analysis is
performed in the framework of the BESIII offline software system which
takes care of the detector calibration and event reconstruction.

The decay $\jpsi \to \etap e^+e^-$ is simulated according to the
Lorentz-invariant amplitude, taking into account the $\jpsi$
polarization state in the $e^+e^-$ annihilation system.  The $\jpsi$
to $\etap$ transition form factor is assumed to be a single-pole form
$|F(q^2)|=1/(1-q^2/\Lambda^2)$, where $q^2$ is the four-momentum
transfer squared and $\Lambda$ is the effective pole mass, with a
value of $\Lambda$ =3.686 GeV/$c^2$.  Then the subsequent decay mode
of $\etap \to \gamma \pi^+\pi^-$ is simulated with the $\rho$-$\omega$
interference and box anomaly effects~\cite{qinlq,qinlq_gampipi}. The
decays of $\etap \to \pi^+\pi^-\eta$ and $\eta \to \gamma\gamma$ are
generated with a phase space model.  The decays of $\jpsi \to \etap V$
($V$ represents $\rho, \omega, \phi$) and $\jpsi \to \etap \ub $ are
generated by a P-wave decay model, and the decay $\ub \to e^+e^-$ is
modeled as a vector meson decaying to a lepton pair~\cite{r_evtgen}.

\section{\boldmath Data analysis}

In this work, the signal is reconstructed with a pair of an electron and
positron, in addition to an $\etap$ meson, which is reconstructed with
two decay modes of $\etap \to \gamma \pi^+ \pi^- $ and
$\etap \to \pi^+ \pi^- \eta(\gamma \gamma)$. The final states of the
corresponding $\jpsi$ decays are $\gamma \pi^+\pi^- e^+e^-$ and
$\gamma\gamma \pi^+\pi^-e^+e^-$, respectively. They are denoted as
mode I and mode II throughout this paper.

Charged tracks in the BESIII detector are reconstructed from hits in
the MDC. We select good charged tracks passing within $\pm$10 cm from
the interaction point (IP) in the beam direction and within 1 cm in
the plane perpendicular to the beam. The polar angle of the track is
required to satisfy $|\cos \theta|<0.93$.  Four candidate charged
tracks are required, and their net charge must be equal to zero.  The
combined information of the energy loss $dE/dx$ from the MDC and the
time of flight from the TOF is used to calculate the PID confidence
levels (C.L.) for the $e,\pi$ and $K$ hypotheses.  Both the electron
and positron selections require the electron hypothesis to have the
highest PID C.L. among the three hypotheses, and the other two charged
tracks are treated as $\pi^+\pi^-$ candidates without any PID
requirement.  The four charged tracks $\pi^+\pi^-e^+e^-$ must pass a
common vertex constrained fit to ensure that they originate from the
interaction point.

Electromagnetic showers are reconstructed from clusters of energy
deposits in the EMC. The shower energy of photon candidates in the EMC
must be greater than 25 MeV in the barrel region
($|\cos \theta| < 0.80$) or 50 MeV in the end-cap region
($0.86<|\cos \theta| < 0.92$).  Showers located between the barrel and
end-cap regions are excluded due to worse reconstruction. Showers are
required to be separated from the extrapolated positions of any
charged track by more than 10$^\circ$. Cluster-timing requirements
are used to suppress electronic noise and unrelated energy
deposits. We require at least one (two) candidate photon(s) for mode I
(II). The $\eta$ meson is reconstructed with $\gamma\gamma$ state,
with the $\gamma\gamma$ invariant mass $M(\gamma\gamma)$ of candidates
required to be within [0.48, 0.60] GeV/$c^2$.

A four-constraint (4C) energy-momentum conservation kinematic fit is
performed to the signal hypothesis. For events with extra photon
candidates, the combination of final state particles with the minimum
chi-square($\chi^2_{\rm 4C}$) is selected, and the $\chi^2_{\rm 4C}$
is required to be less than 100. The $\chi^2_{\rm 4C}$ requirement
removes more than 12\% and 30\% of background events for mode I and II,
respectively, and results in a signal efficiency loss of about 7\% for
both modes.

\subsection{\boldmath Measurement of $\mathcal{B}(\jpsi \to \etap
  e^+e^-$)}
\label{sec_bf}

\begin{figure}[!htb]
\centering
\includegraphics[width=0.24\textwidth]{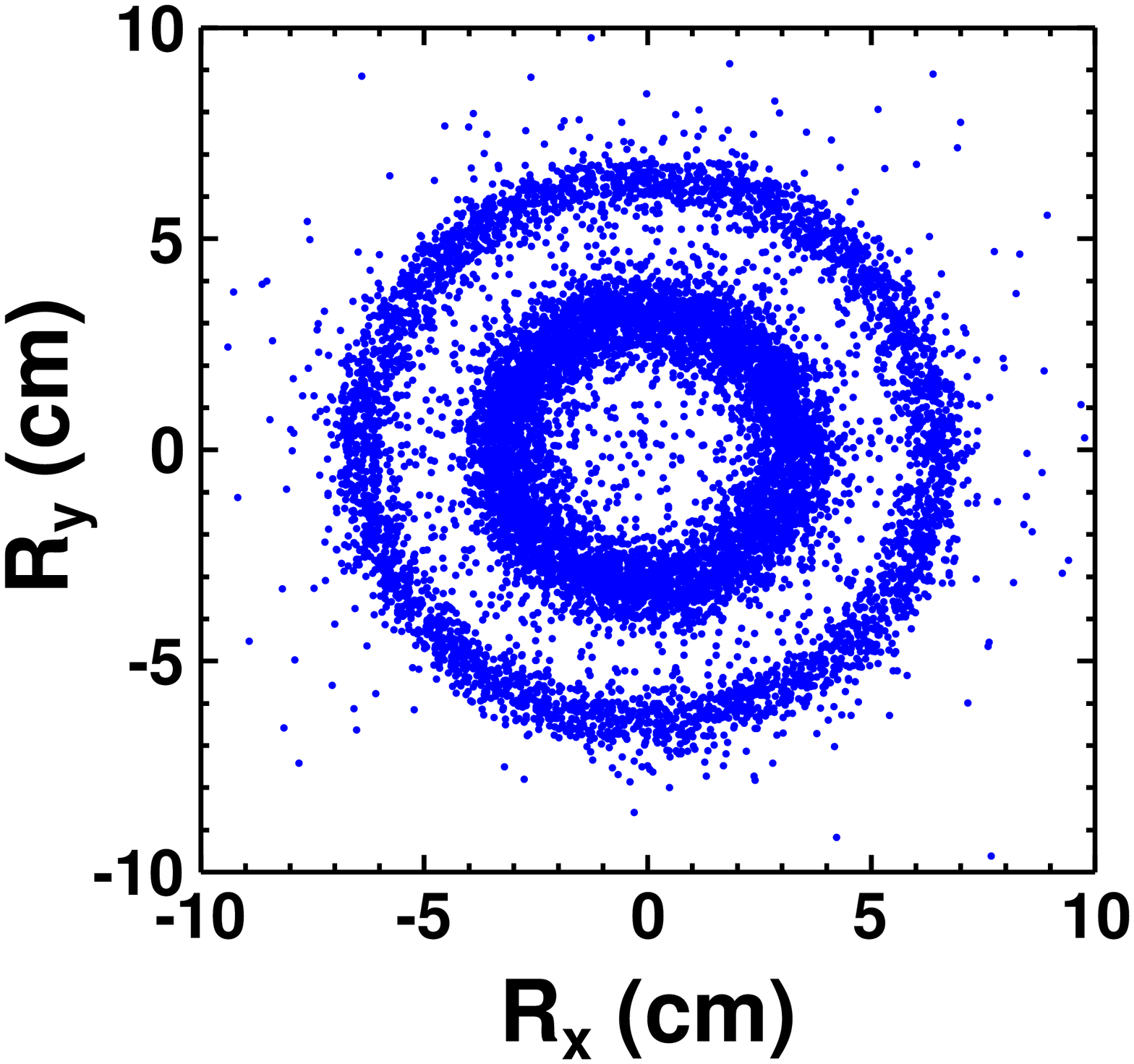}\put(-20,90){\bf ~(a)}
\includegraphics[width=0.24\textwidth]{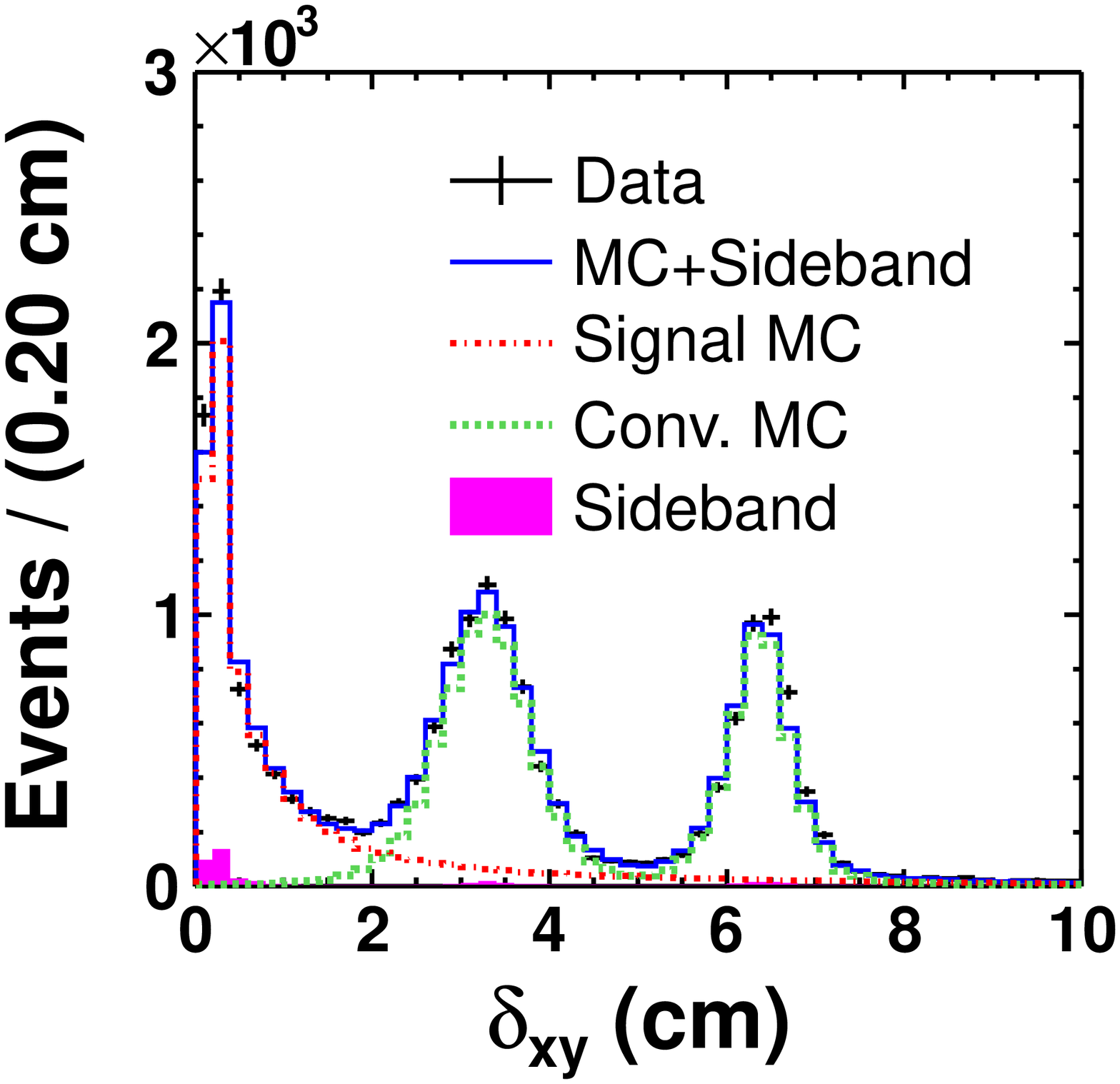}\put(-20,90){\bf ~(b)}
\caption{  Electron-positron pair vertex
  distributions. (a) The two-dimensional scatter plot of $R_y$ versus
  $R_x$ from $\jpsi\to \gamma\etap(\etap \to \gamma \pi^+\pi^-)$ MC
  simulated events. (b) Distribution of $\delta_{xy}$. The black
  crosses are data, the blue solid line is the total contribution from
  the MC and $\etap$ sideband, the green dotted-dashed line is signal MC, the
  dashed line is $\jpsi\to \gamma\etap(\etap \to \gamma \pi^+\pi^-)$
  MC and the shaded area is the  $\etap$ sideband. The quantities are
  described in the text. }
\label{f_rxy}
\end{figure}

The major background that can peak in $\etap$ mass distribution is
from the radiative decay $\jpsi \to \gamma \etap$ followed by a
$\gamma$ conversion process, where $\gamma$ converts into an $e^+e^-$
pair when it interacts with material in front of the MDC.  The
distance from the reconstructed vertex of the $e^+e^-$ pair to the IP
in the $x$-$y$ projection, $\delta_{xy}$, is used to identify $\gamma$
conversion~\cite{r_conv}.  Here $\delta_{xy}=\sqrt{R_x^2+R_y^2}$, and
$R_x$ and $R_y$ are the coordinates of the reconstructed vertex in the
$x$ and $y$ directions. The scatter plot of $R_y$ versus $R_x$ of the
$\jpsi \to \gamma \etap (\etap \to \gamma \pi^+\pi^-)$ MC sample, and
the $\delta_{xy}$ distributions of data and various MC samples are
shown in Fig.~\ref{f_rxy}. The two peaks above 2 cm in the
$\delta_{xy}$ distribution match the positions of the beam pipe and
inner wall of the MDC. Only events with $\delta_{xy} < 2$ cm are
retained. The normalized number of the remaining $\gamma$ conversion
events is estimated according to the corresponding BFs
from the PDG~\cite{r_pdg16}, as $202.2 \pm 7.3$ ($70.6 \pm 2.5$) in
mode I (II).

In addition to the $\gamma$ conversion events, there are some other
minor backgrounds that can also peak in $\etap$ mass distribution. The
background from $\jpsi \to V \etap, V = \rho, \omega,\phi$ decays is
studied with high statistics MC samples.  The peaking background from
the process $\jpsi \to \etap \pi^+\pi^-$ is estimated with a MC sample
generated according to the amplitude as reported in
Ref.~\cite{r_lhj_etappipi}.  The numbers of these background events
($N^{\rm bkg}$), normalized according to the world-averaged
BFs~\cite{r_pdg16}, are summarized in Table~\ref{t_peakbkg}. Potential
peaking background from the two-photon process $e^+e^-\to e^+e^-\etap$
is found to be negligible, as studied with 2.93 fb$^{-1}$ of data at
c.m. energy of 3.773 GeV~\cite{lumi3770,pipics}.

\begin{table}[!htb]
  \caption{Number of nonconversion peaking background events, as
    estimated with high statistics MC. The uncertainties include those
    of all intermediate resonance decay BFs~\cite{r_pdg16}. ``--''
    indicates cases of no event survival. The first group lists
    contributions from $\jpsi \to \phi \etap(\phi \to e^+e^-)$, as
    fixed in the fitting, and
    $\jpsi \to \rho \etap(\rho \to \pi^+\pi^-)$, which is included
    coherently in $\jpsi \to \etap \pi^+\pi^-$. The second group shows
    minor contribution sources, which are not accounted for in the
    fitting, but needs to be subtracted from the fitted signal yield.}
\begin{center}
\begin{tabular}{c|c|c}
\hline
\hline
& $N^{\rm bkg}$ (mode I) & $N^{\rm bkg}$ (mode II) \\
\hline
$\jpsi \to \phi \etap(\phi \to e^+e^-) $    &$17.1\pm1.9$  & $6.4\pm0.7$  \\
$\jpsi \to \rho \etap(\rho \to \pi^+\pi^-)$ &$2.8\pm0.2$   & $0.8\pm0.1$ \\
% $\jpsi \to \rho \etap(\rho \to \pi^+\pi^-)$ \footnotemark[1] &$2.8\pm0.2$   &
%                                                             $0.8\pm0.1$ \\

\hline
$\jpsi \to \omega \etap(\omega \to e^+e^-)$   &$1.6\pm0.2$   & $0.6\pm0.1$ \\
$\jpsi \to \rho \etap(\rho \to e^+e^-)$     &$0.48\pm0.05$   & $0.18\pm0.02$ \\
$\jpsi \to \phi \etap(\phi \to K^+ K^-)$ & --   & --          \\
$\jpsi \to \etap \pi^+\pi^-$ &$ 3.8 \pm 0.2$   & $1.8 \pm 0.1$ \\
\hline
Contribution to subtract & $5.9\pm0.3$  & $2.5\pm0.2$  \\
\hline
\hline
\end{tabular}
\end{center}
%\footnotetext[1]{Not included in the total to avoid double counting with $\jpsi \to \etap \pi^+\pi^-$ }
\label{t_peakbkg}
\end{table}

Mode II has no obvious nonpeaking contamination of the $\etap$
invariant mass distribution, and the background containing an $\eta$ is
determined to be negligible from a MC study.  There are two kinds of
nonpeaking background for mode I. One is
$\jpsi \to \pi^+ \pi^- \eta/\pi^0$ with
$ \eta/\pi^0 \to \gamma e^+e^-$, which has the same final state
$\gamma \pi^+ \pi^- e^+e^-$ as the signal process.  To reject the
background with a $\pi^0$ intermediate state, candidates with the
$\gamma e^+e^-$ invariant mass $M(\gamma e^+e^-)$ in the $\pi^0$ mass
window [0.10, 0.16] GeV/$c^2$ are removed. The other is from $\jpsi$
decays with multiple pions in the final state, where a pion pair is
misidentified as an electron-positron pair. Both backgrounds produce a
smooth shape on the $\gamma \pi^+ \pi^-$ invariant mass
$M(\gamma \pi^+\pi^-)$ distribution around the $\etap$ mass.

\begin{figure}[!htb]
\centering
\includegraphics[width=0.24\textwidth]{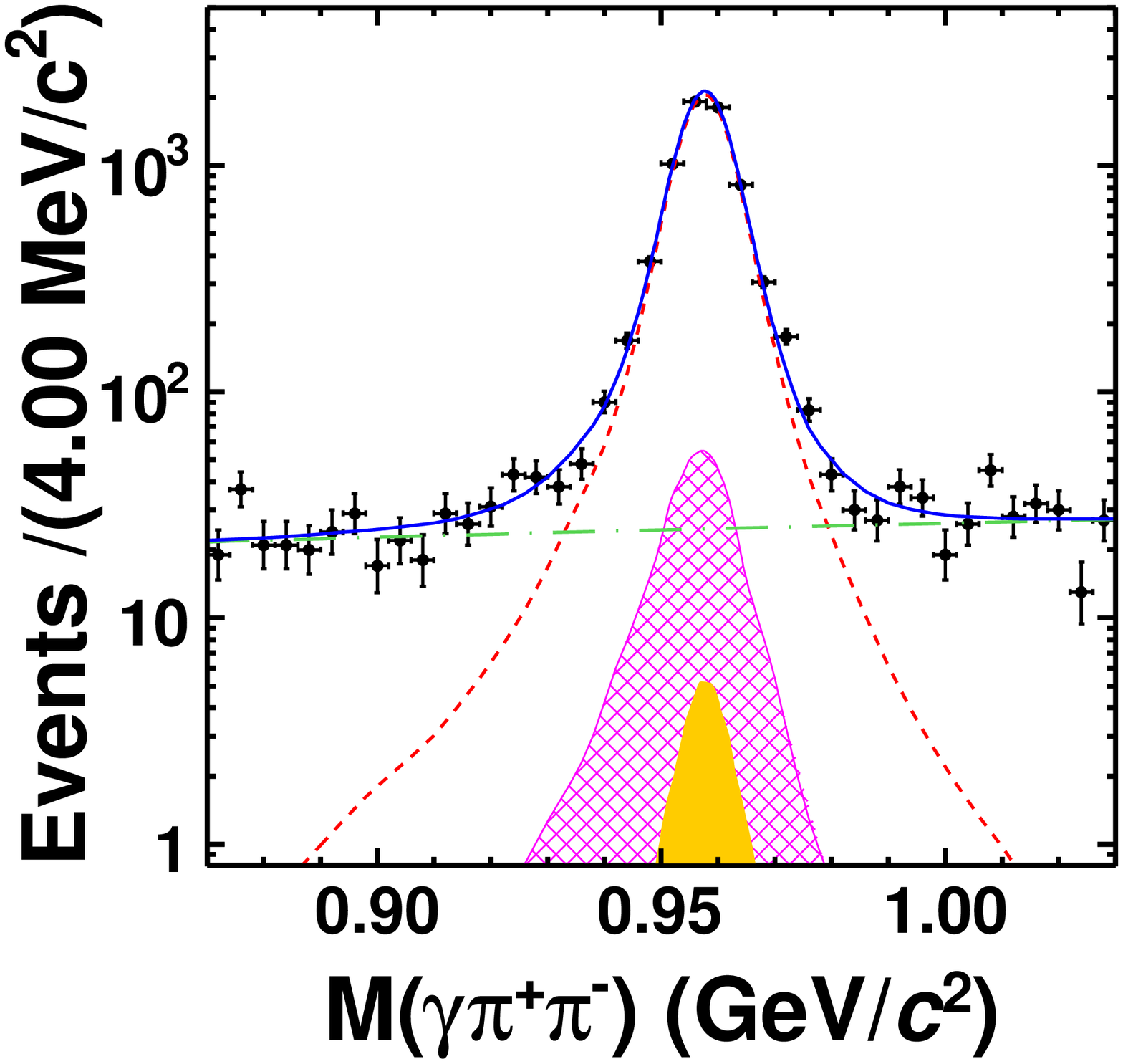}\put(-25,90){\bf ~(a)}
\includegraphics[width=0.24\textwidth]{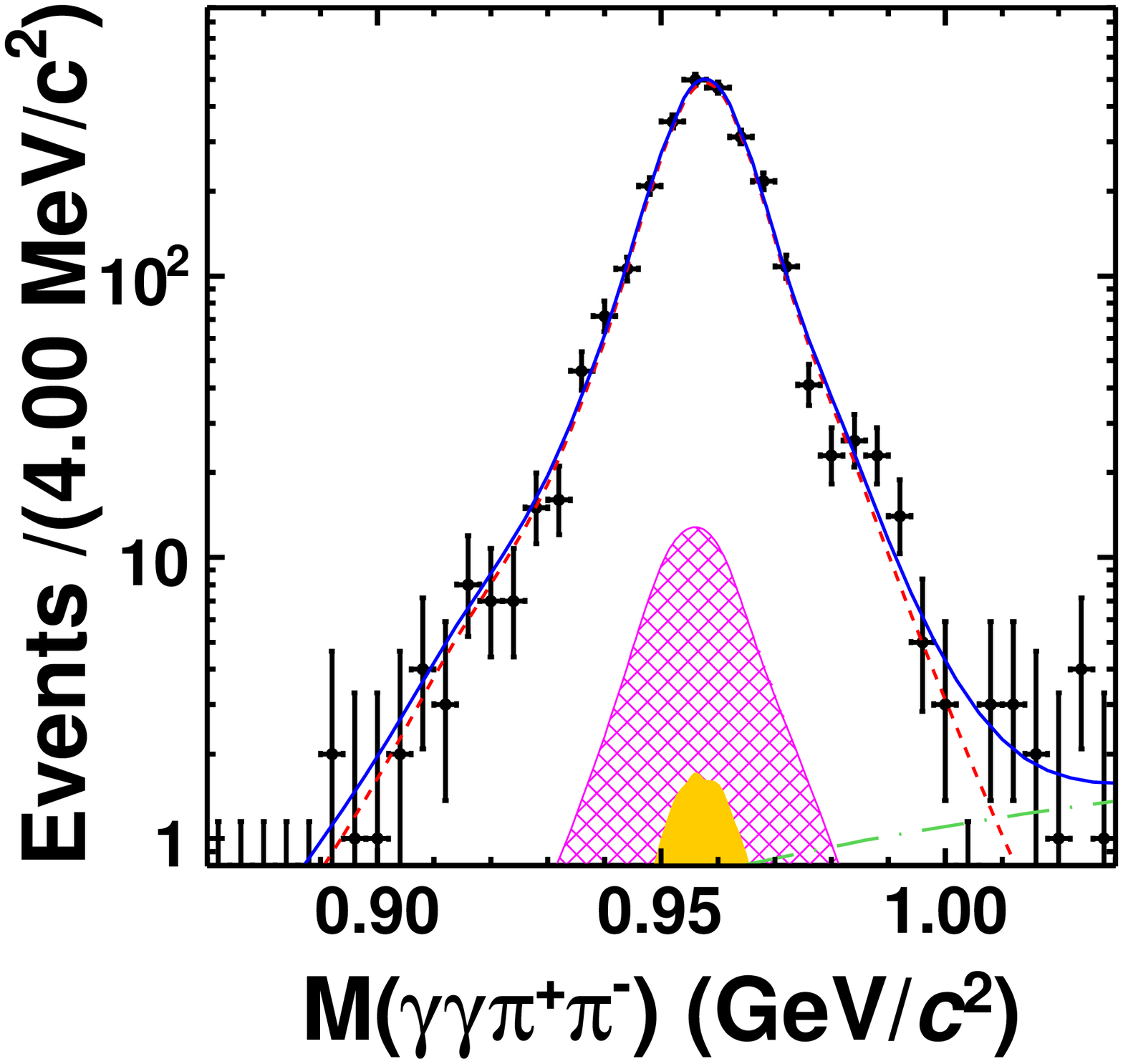}\put(-25,90){\bf ~(b)}
\caption{ The invariant mass spectra of $\etap$
  candidate events (a) $M( \gamma \pi^+ \pi^-)$ and (b)
  $M(\gamma \gamma \pi^+\pi^-)$. Black crosses are data, the blue
  solid line is the total fitting projection, the red dashed line is
  signal, the green dotted-dashed line is nonpeaking background, the
  pink cross-hatched area is $\gamma$ conversion
  $\jpsi \to \gamma \etap$ and the yellow solid area is
  $\jpsi \to \etap \phi, \phi \to e^+e^-$.}
\label{f_fitbr}
\end{figure}

The distributions of $\gamma \pi^+ \pi^-$ invariant mass
$M(\gamma\pi^+\pi^-)$ and $\gamma\gamma \pi^+\pi^-$ invariant mass
$M(\gamma\gamma\pi^+\pi^-)$ of surviving candidate events after all
the above selection criteria, within the region [0.87, 1.03]
GeV/c$^2$, are shown in Fig.~\ref{f_fitbr}.

Unbinned maximum likelihood (ML) fits are performed on the
$M(\gamma \pi^+\pi^-)$ and $M(\gamma\gamma \pi^+\pi^-)$ distributions
to determine the signal yields. In the fits, the signal probability
density function (PDF) is described by a signal MC simulated shape
convolved with a Gaussian function, which takes into account the
resolution difference between data and MC simulation. The major
peaking backgrounds from $\gamma$ conversion $\jpsi \to \gamma \etap$
and $\jpsi \to \phi \etap(\phi \to e^+e^-)$ are described with MC
shapes, and their magnitudes are fixed to the expected values. The
number of minor peaking background events as shown in
Table~\ref{t_peakbkg} is directly subtracted from the fitted $\etap$
yields. The nonpeaking backgrounds in mode I and II are described
with second- and first-order Chebyshev polynomial functions,
respectively. The fit results are shown in Fig.~\ref{f_fitbr}, with
signal yields of $6442.8 \pm 87.1$ and $2497.9 \pm 51.3$ for mode I
and II, respectively. The goodness of fit is demonstrated by $\chi^2$
over the number of degrees of freedom (ndf), with values $\chi^2/ndf$ =
74.8/35 and 34.3/17 for mode I and II, respectively.  The BF of
$\jpsi \to \etap e^+e^-$ is determined by

\begin{equation}
\label{eq_bf}
\mathcal{B}(\jpsi \to \etap e^+e^-)=\frac{N_{\rm sig}}{N_{\jpsi} \cdot
  \mathcal{B}_{\etap \to {\rm F}} \cdot \mathscr{E}\cdot \delta^2},
\end{equation}
where $N_{\rm sig}$ is the number of signal events, $N_{\jpsi}$ is the
number of $\jpsi$ events, $\mathcal{B}_{\etap \to {\rm F}}$ is the
intermediate BF of the $\etap$ decay, ($28.9 \pm 0.5$)\% for
$\mathcal{B}(\etap\to\gamma\pi^+\pi^-)$ in mode I and
($16.9 \pm 0.3$)\% for
$\mathcal{B}(\etap\to\eta \pi^+ \pi^-) \times \mathcal{B}(\eta \to
\gamma \gamma)$ in mode II~\cite{r_pdg16}, $\mathscr{E}$ is the
detection efficiency ($25.01\pm0.06$)\% for mode I and
($16.82 \pm 0.06$)\% for mode II and $\delta=1.012$ is the tracking
efficiency correction factor per electron/positron as described in
Sec.~\ref{sec_syst}.  Using Eq.~(\ref{eq_bf}) and taking into account
the systematic uncertainties discussed in Sec.~\ref{sec_syst},
$\mathcal{B}(\jpsi \to \etap e^+e^-)$ values for mode I and mode II are
calculated to be $(6.63 \pm 0.09 \pm 0.21 )\times 10^{-5}$ and
$(6.54 \pm 0.13 \pm 0.26 )\times 10^{-5}$, respectively, where the
first uncertainties are statistical and the second are systematic. The
results from the two $\etap$ decay modes are consistent with each
other within the statistical and uncorrelated systematic
uncertainties. With the weighted least squares method taking into
account the correlated and uncorrelated
uncertainties~\cite{r_nimchi2}, the weighted average BF from the two
decay modes is
$\mathcal{B}(\jpsi \to \etap e^+e^-) = (6.59\pm0.07\pm0.17)\times
10^{-5}$. This result is consistent with the previous BESIII
measurement~\cite{r_chuxkpee}, and the precision is improved by a
factor of 2, from 6\% to 3\%. The measured value is higher than the
theoretical prediction of Ref.~\cite{r_fjl} from the single-pole form
factor and will provide further input to improve theoretical models.

\subsection{\boldmath Search for the dark photon through $\jpsi \to \etap e^+e^-$ }
The dark photon is searched for  by looking for a narrow resonance peaking
on a smooth electron-positron invariant mass [$M(e^+e^-)$] spectrum at
the position of the dark photon mass ($m_{\ub}$). Candidate events
with the $\etap e^+e^-$ final state are selected with the same
selection criteria as described in Sec.~\ref{sec_bf} but without the
$\gamma$ conversion veto criteria. Since $\gamma$ conversion events
distribute mainly in the low $M(e^+e^-)$ region below 70 MeV/$c^2$,
only candidate events with $M(e^+e^-)>$ 70 MeV/$c^2$ are retained.
The mass ranges [0.74, 0.84] GeV/$c^2$ and [1.00, 1.04] GeV/$c^2$,
corresponding to the regions of the $\omega$ and $\phi$ mesons, are
excluded in the dark photon search, since the search sensitivity of
$\ub \to e^+ e^-$ would degrade significantly due to the complicated
SM background and the suppressed BF of
$\ub \to e^+ e^-$~\cite{Batell2009, wang_gev, Li2010}. The invariant
mass of the $\etap$ candidates is required to be within [0.93, 0.98]
GeV/$c^2$ and the selected events are mainly from the EM Dalitz decay
$\jpsi \to \etap e^+e^-$.

% In the region of the $\omega$ and $\phi$ mesons, the BF of
% $\ub \to e^+ e^-$ is greatly suppressed due to the enhancement of
% $\ub$ decays into hadronic channels~\cite{Batell2009, wang_gev,
%   Li2010} and these resonances complicates the SM background in their
% vicinity, so the search sensitivity of $\ub \to e^+ e^-$ would degrade
% significantly for the $\ub$ mass within these regions. Therefore, the
% mass ranges [0.74, 0.84] GeV/$c^2$ and [1.00, 1.04] GeV/$c^2$,
% corresponding to the regions of the $\omega$ and $\phi$ mesons,
% respectively, are excluded in the dark photon search.

\begin{figure}[!htb]
\includegraphics[width=0.24\textwidth]{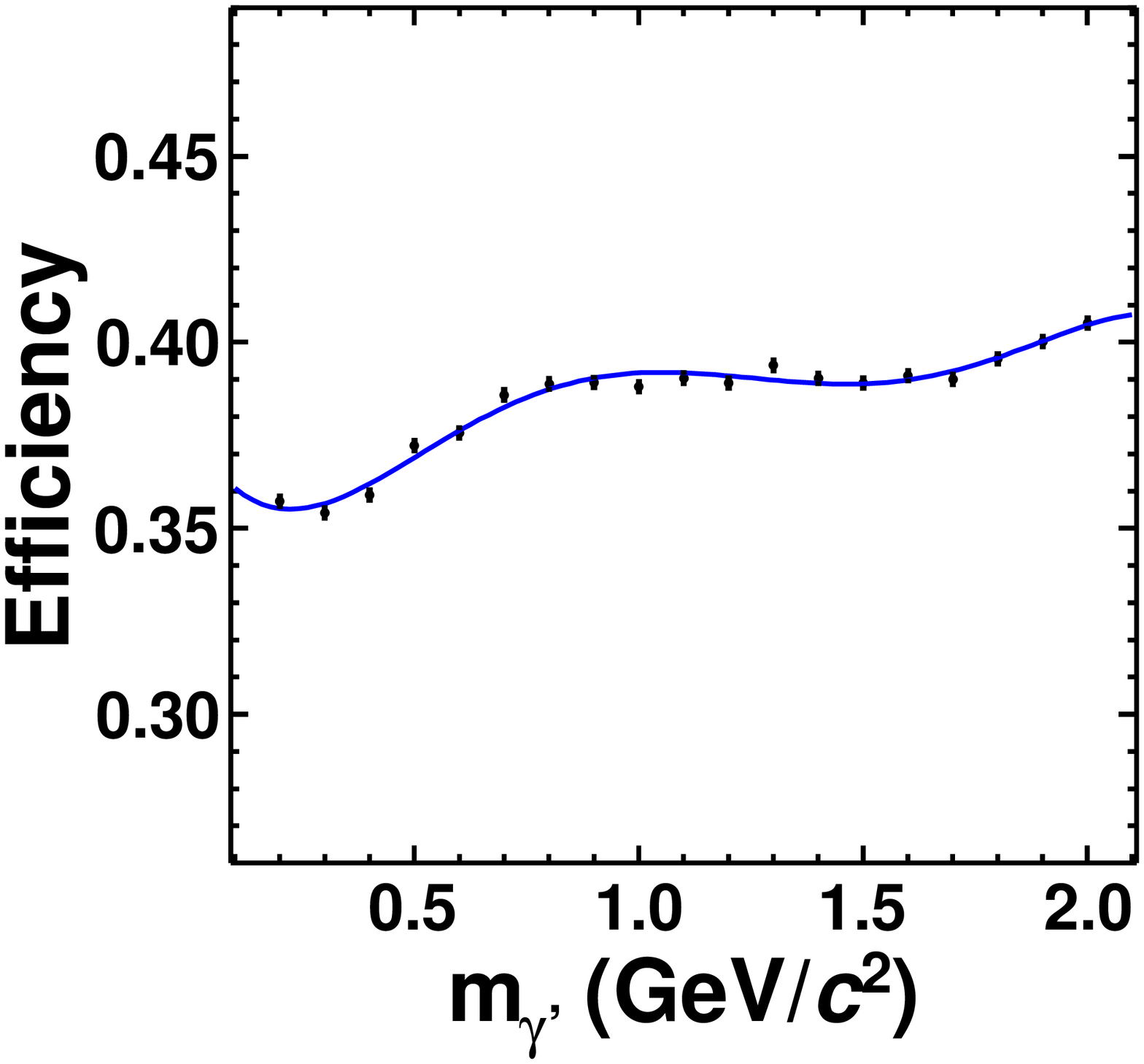}\put(-25,95){\bf ~(a)}
\includegraphics[width=0.24\textwidth]{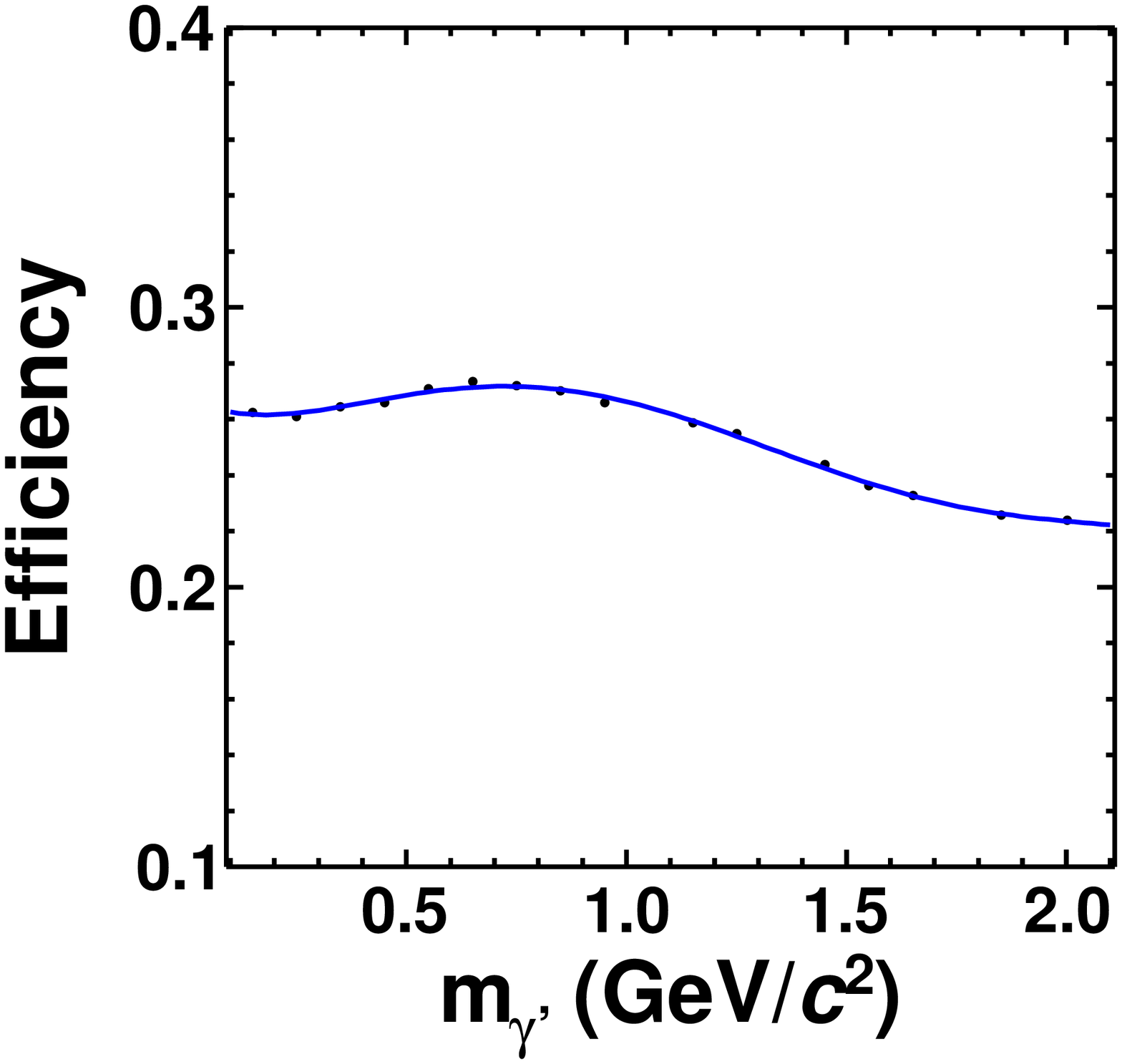}\put(-25,95){\bf ~(b)}
\caption{ The total detection efficiency of the entire decay chain for
  different $m_{\ub}$ of (a) mode I and (b) mode II. The blue
  curves show the fit with polynomial functions.  }
\label{f_signal}
\end{figure}

\begin{figure*}[!htb]
\centering
\subfigure{
\includegraphics[width=0.45\textwidth]{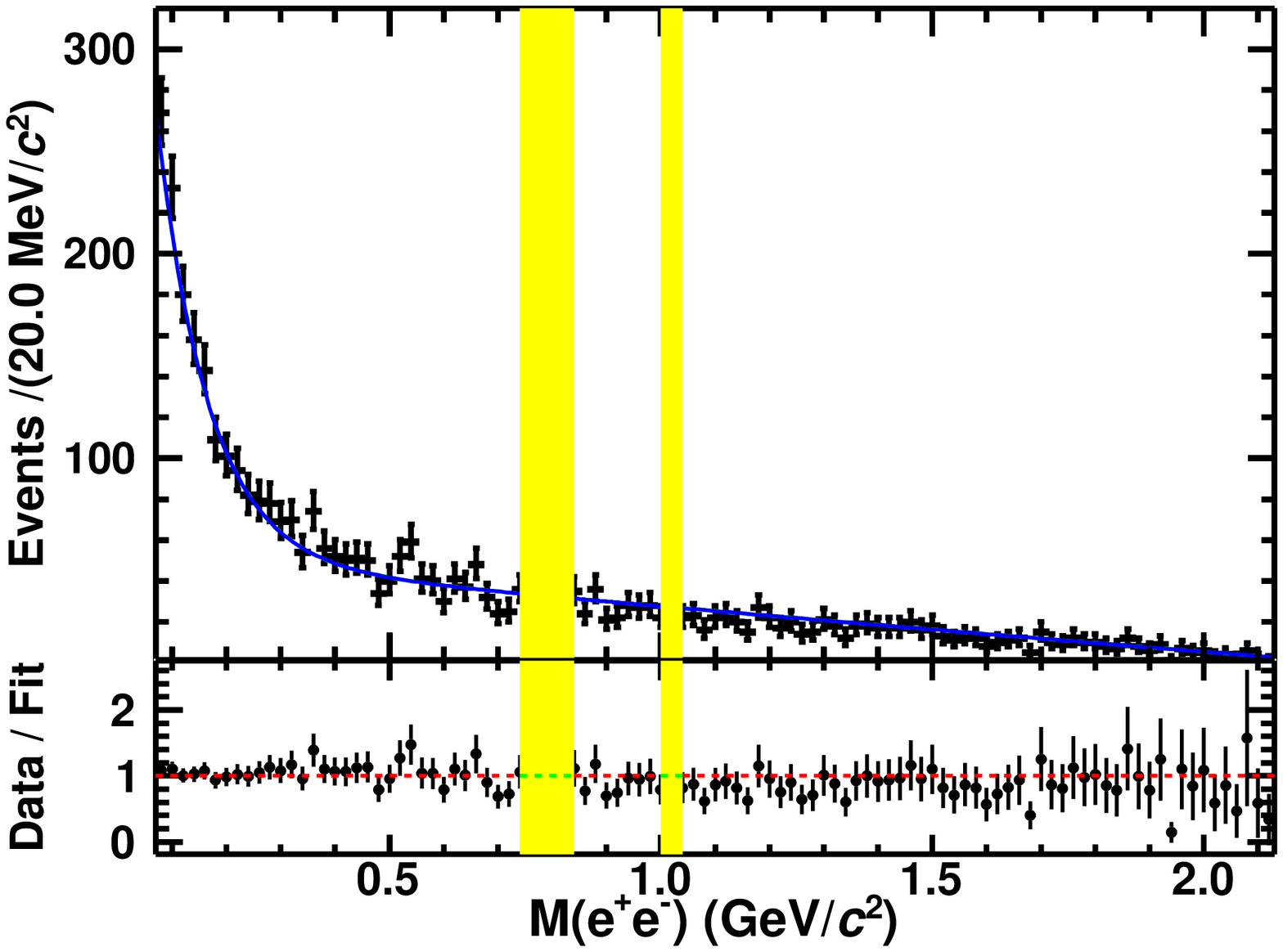}}\put(-30,130){\bf ~(a)}
\subfigure{
\includegraphics[width=0.45\textwidth]{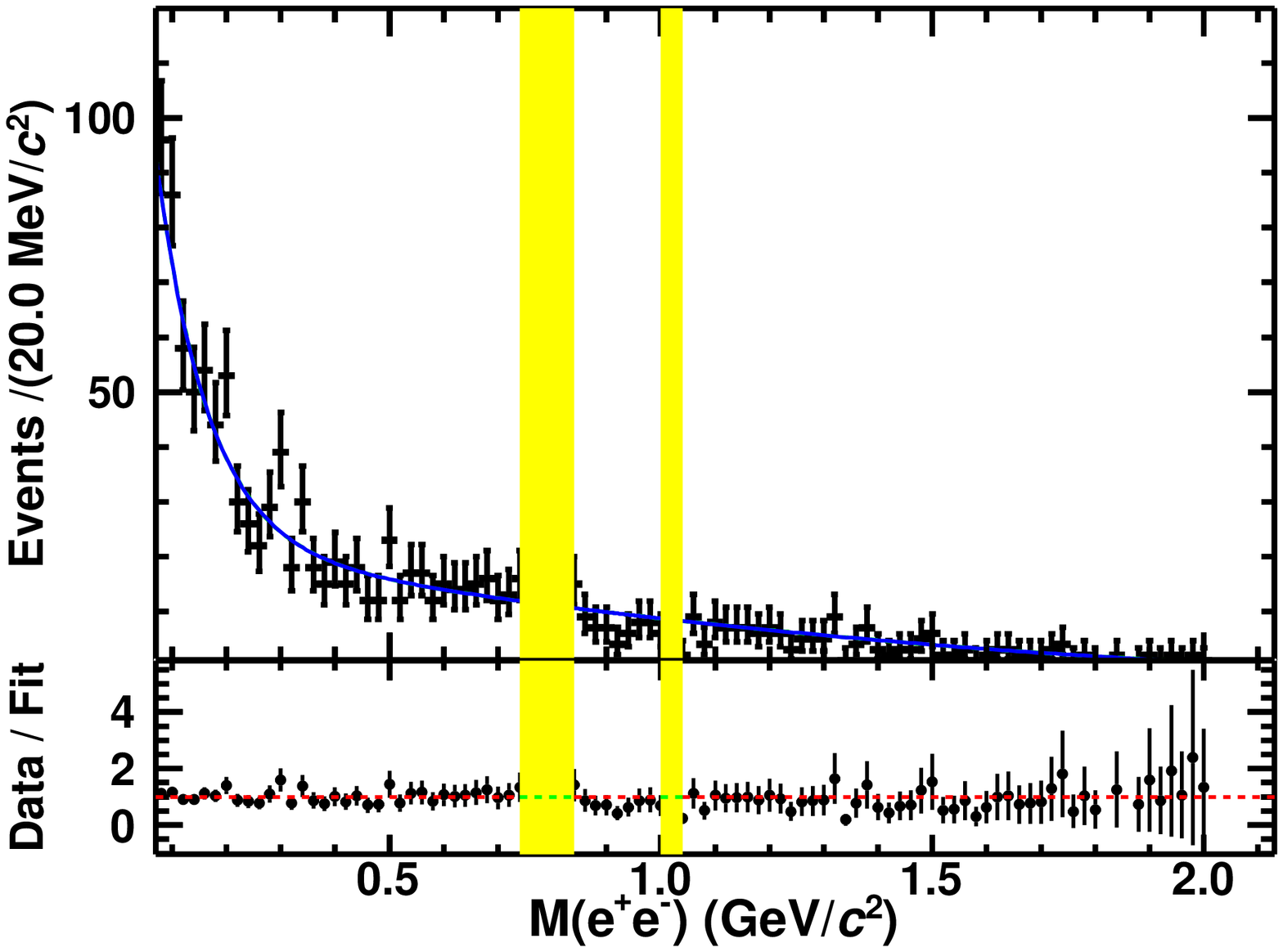}}\put(-30,130){\bf ~(b)}{}
\subfigure{
\includegraphics[width=0.45\textwidth]{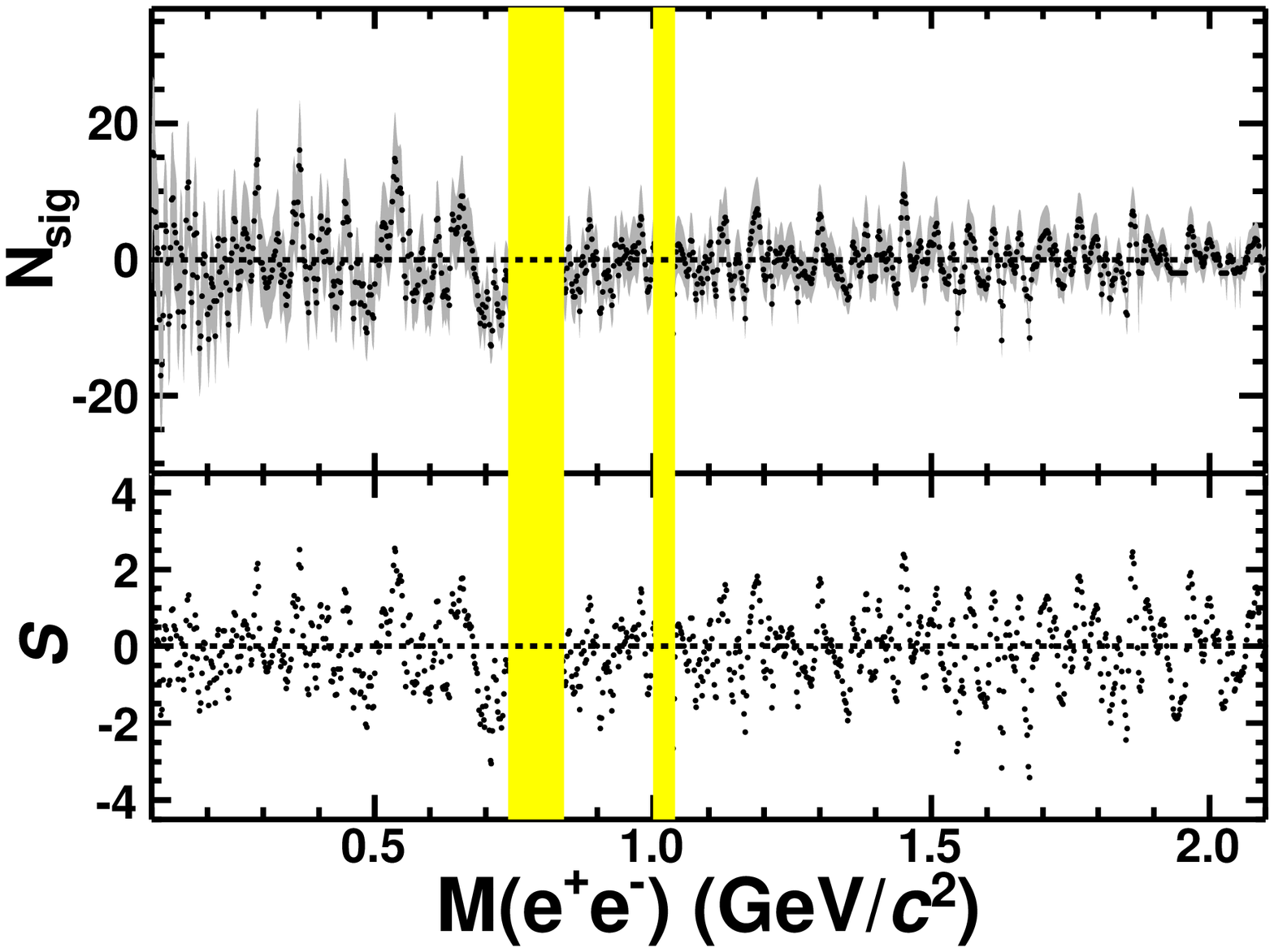}}\put(-30,130){\bf ~(c)}
\subfigure{
\includegraphics[width=0.45\textwidth]{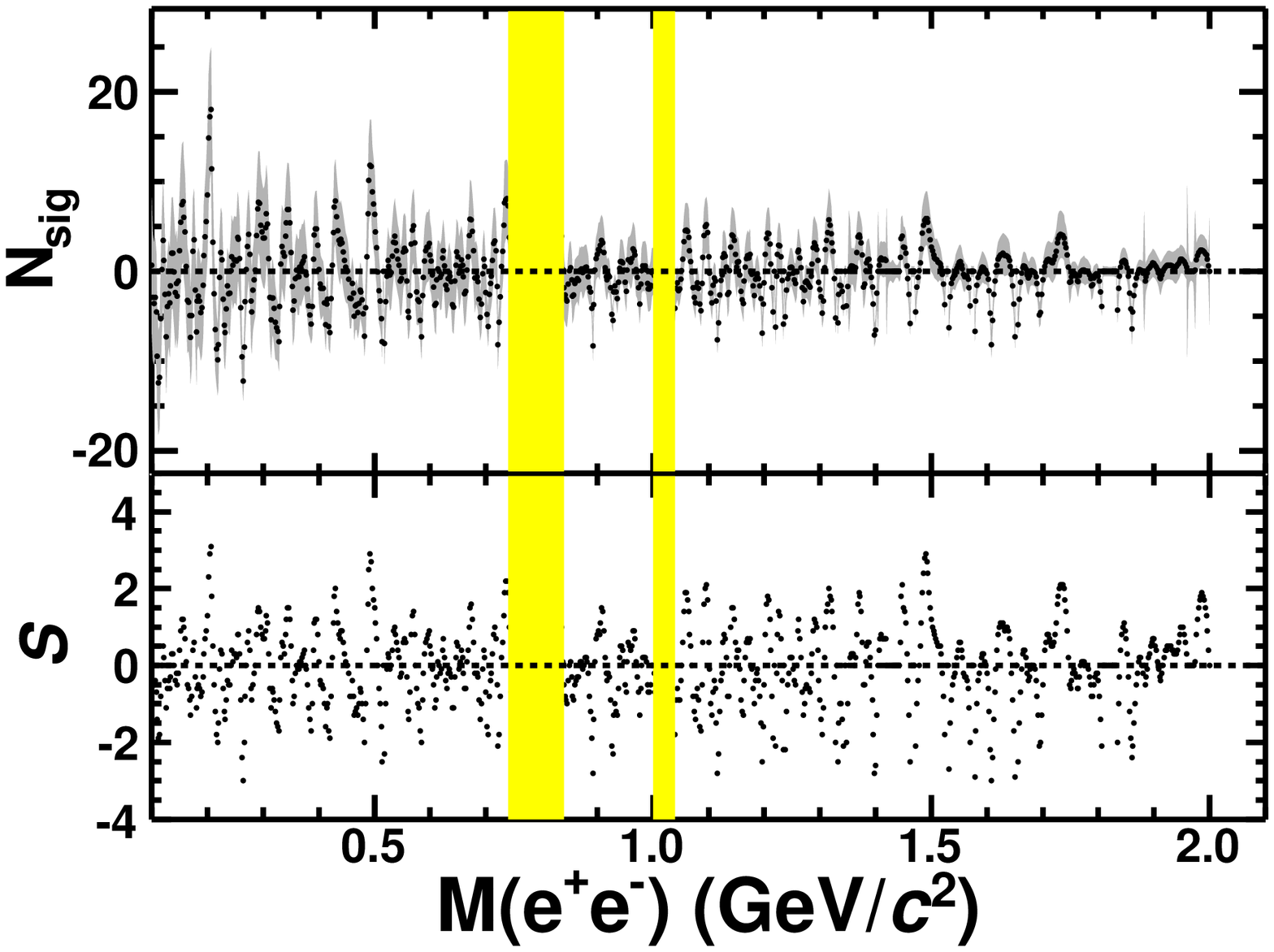}}\put(-30,130){\bf ~(d)}{}
\caption{ The upper plots in (a) and (b) show the electron-positron
  pair invariant mass distributions for mode I and mode II,
  respectively, where the crosses are data, and the solid lines are
  the background-only fitting results. The lower plots in (a) and (b)
  show the ratio of data over fitting result for each bin. The shaded
  bands are the corresponding excluded $\omega$ and $\phi$
  regions. The plots in (c) and (d) are the number of signal
  events ($N_{\rm sig}$) and the significances ({\it S}) from the fit
  for mode I and mode II, respectively. }
\label{f_mee}
\end{figure*}

The signal PDF and detection efficiency are determined with a series of
signal MC samples. They are generated according to the decay chain
$\jpsi \to \etap \ub, \ub \to e^+e^-$, with different $m_{\ub}$
values, ranging from 0.1 to 2.0 GeV/$c^2$ with a step of 0.1
GeV/$c^2$. The dark photon width, suppressed by a factor of
$\varepsilon ^2$ and expected to be far below the experimental
resolution, is set to zero in the MC generation.  The dark photon
signal PDF is parametrized by the sum of two Crystal Ball (CB)
functions with a common mean value, where the parameters are
determined by fitting the signal MC samples. The resolution, which is
evaluated by weighting the widths of two CBs according to their ratio,
grows from 2 to 8 MeV/$c^2$ as $m_{\ub}$ increases.  The
detection efficiency, shown in Fig.~\ref{f_signal}, ranges from 35\%
to 41\% and from 22\% to 27\% depending on $m_{\ub}$ in mode I and II,
respectively. The efficiency and signal PDF parameters are
interpolated between the mass points by a fit with a polynomial
function. The background PDF is the sum of a second-order polynomial
function and an exponential function:
$f(\mee)=c_2\cdot\mee^2+c_1\cdot \mee+c_0+e^{c_3\cdot\mee}$. The
parameters $c_0, c_1,c_2, and c_3$ are determined from a background-only ML
fit of data as shown in Figs.~\ref{f_mee}(a) and (b).

To determine the possible dark photon signal yield, a series of ML fits are
performed in the range 0.07 $<m_{\ub}<$ 2.13 GeV$/c^{2}$ with uniform
mass steps of 2 MeV/$c^{2}$. In each fit, a composite PDF model of the
corresponding signal shape and the common background description is
used, with the parameters of the signal and background fixed while
their yields are free to float. In order to avoid fit failure due to
the limited statistics in the high $M(e^+e^-)$ region in mode II, a
lower bound on the signal yield is imposed by requiring the total PDF
of signal plus background to remain non-negative.

For each $M(e^+e^-)$ point, the local significance of the signal is
determined by
$\mathcal{S} = {\rm sign}(N_{\rm sig}) \sqrt{-2{\rm
    ln}(\mathcal{L}_0/\mathcal{L}_{\rm max})}$, where $\mathcal{L}_0$
($\mathcal{L}_{\rm max}$) is the likelihood value without (with) the
signal hypothesis included in the fit.  The results of $N_{\rm sig}$
and the corresponding significance are shown in Figs.~\ref{f_mee}(c)
and (d), respectively. The maximum local significance is from mode II,
with $3.1\sigma$ at 0.204 GeV/$c^2$. The corresponding global
significance is less than 1$\sigma$, evaluated by using a large number
of pseudoexperiments~\cite{r_vindy180}. In conclusion, no significant
dark photon signal is observed within the searched range.

\section{\boldmath Systematic uncertainties}
\label{sec_syst}
Most of the systematic uncertainties from the event selection are the
same for the BF measurement and the dark photon search. The
correlations between the two $\etap$ reconstruction modes are taken
into account when evaluating the uncertainties for the BF
measurement. The uncertainties for the efficiencies of MDC tracking,
photon detection, PID, the number of $J/\psi$ decay
events, and the $\gamma$ conversion veto are considered as correlated
sources. Those for an additional photon in
$\etap \to \pi^+\pi^- \eta(\gamma \gamma)$, the 4C kinematic fit,
$\eta$ reconstruction, the form factor, signal shape, fit range,
background shape and magnitude and $\etap$ BFs are considered as the
independent sources.  The systematic uncertainties are discussed
below and summarized in Table~\ref{t_sys}.

Some of the uncertainties are estimated in a similar way as described
in Ref.~\cite{r_chuxkpee}. The uncertainty is 0.6\% per electron due
to PID, determined by comparing the efficiency difference between data
and MC simulation for a control sample of radiative Bhabha
$e^+e^-\to\gamma e^+e^-$ (including $\jpsi \to\gamma e^+e^-$) events
collected at the $\jpsi$ energy.  With the same control sample, the
electron tracking efficiency from MC is corrected in a two-dimensional
distribution of the transverse momentum versus polar angle of the
lepton tracks by different interpolation algorithms event by
event. The difference between data and MC after correction shows
consistent values of 0.5\% per track, which is taken as the electron
tracking uncertainty and the overall efficiency correction factor per
electron/positron track is calculated to be $\delta=1.012$. The
charged pion tracking efficiency is studied with the control sample of
$\jpsi \to \pi^+ \pi^- p \bar{p}$ events.  The differences between data
and MC are tabulated in bins of transverse momentum and polar
angle. After reweighting according to the signal kinematics, the
tracking uncertainties for charged pions are determined to be 0.3\%
per pion track in mode I and 0.7\% in mode II, reflecting the
different pion transverse momentum distributions of the two
modes. Tracking uncertainties are treated as fully correlated and thus
added linearly.

The  photon detection efficiency is studied with a control sample
of $\jpsi \to \pi^+\pi^-\pi^0, \pi^0 \to \gamma \gamma$ events. The
data/MC difference is 0.5\% (1.5\%) for a photon in the EMC barrel
(end-cap) region. The average difference, 0.6\% per photon, is taken
as systematic uncertainty.

The uncertainty associated with the 4C kinematic fit is estimated with a high
purity control sample of $\jpsi \to \pi^+\pi^-\pi^0, \pi^0 \to
\gamma e^+e^-$ events. The efficiency difference between data and MC simulation is
0.5\%, which is taken as the systematic uncertainty. This control
sample is also used to estimate the uncertainty due to the $\gamma$
conversion veto criterion $\delta_{xy} < 2$ cm. The difference in
efficiency between data and MC simulation is 1\% and is taken as the
uncertainty.

The signal efficiency may also be biased due to the $\eta$
reconstruction via its $\gamma\gamma$ decay. The systematic
uncertainty for it is determined to be 1.0\% from a study of the
control sample of $\jpsi \to p\overline{p}\eta$~\cite{r_etasys}
events.

The uncertainty of the transition form factor used in the MC generation
is estimated with the alternative signal MC samples generated with the
parameters $\Lambda$=3.0 or 4.0 GeV/$c^2$, and the largest efficiency
difference 0.4\% (0.2\%)with respect to the nominal one is taken as
the uncertainty for mode I (II).

The uncertainty on the efficiency due to the choice of the signal
parametrization is 0.4\% (0.2\%) for mode I (II), evaluated by
comparing the signal yields with and without the Gaussian function
convolution in the fit. We select alternative fit ranges and a higher
order Chebyshev polynomial function for nonpeaking background shapes
to estimate the related uncertainty. The largest difference of the
signal yield with respect to the nominal one, 0.6\% (1.3\%), is taken
as the uncertainty for mode I (II). The uncertainty due to fixing the
peaking-background yield is 0.3\% for both modes, evaluated by
adjusting the number of peaking background events by one standard
deviation of the total peaking background yield.

The uncertainty of the number of $\jpsi$ events is determined to be
0.5\%~\cite{r_jpsi_no} and those of the $\etap$ BFs are taken as 1.7\%
for both modes~\cite{r_pdg16}.

For the dark photon search, the systematic uncertainties are divided
into additive and multiplicative terms. The additive systematic
uncertainties arise from the fit bias and the signal and background
PDFs. The multiplicative uncertainties come from the number of
$J/\psi$ events, $\etap$ BFs and detection efficiencies, which have
been discussed in the BF measurement.  To incorporate these
uncertainties, we take the additive systematic uncertainty into
consideration by performing the same fit procedure with different
combinations of the nominal and alternative fit ranges, signal shapes
and background shapes. The maximum number of signal events among the
different fit scenarios is adopted to calculate the the upper limit of
the signal yield $N_{\rm sig}$.  This procedure is performed for
mode I and mode II separately. The multiplicative systematic
uncertainties in the search for a dark photon are listed in
Table~\ref{t_sys}. Most of them come from differences in the selection
efficiency between data and MC simulation. When deriving
$\mathcal{B}(\jpsi \to \etap \ub)$ from the product BF, an additional
systematic uncertainty originates from the theoretical BF of
$\mathcal{B}(\ub \to e^+e^-)$, which is $0\sim 14$\% depending on
$m_{\ub}$ according to Ref.~\cite{Batell2009} and mainly comes
from the R value measurement.

\begin{table*}[!htbp]
 \caption{Sources of systematic uncertainties for the BF measurement
   and multiplicative terms for the dark photon search (in \%).  The
   correlated sources between $\etap \to \gamma \pi^+\pi^-$ and
   $ \etap \to \pi^+ \pi^- \eta(\gamma \gamma)$ modes are marked with
   an asterisk.  }
\begin{center}
\begin{tabular}{p{ 0.36\textwidth}<{\centering} | p{0.15\textwidth}<{\centering} p{0.15\textwidth}<{\centering} | p{0.15\textwidth}<{\centering} p{0.15\textwidth}<{\centering} }
\hline
\hline
\centering
\multirow{2}{*}{Sources}& \multicolumn{2}{c|}{BF measurement}  & \multicolumn{2}{c}{Search for dark photon} \\
\cline{2-5}
& $ \etap \to \gamma \pi^+ \pi^-$   & $ \etap \to \pi^+ \pi^- \eta(\gamma \gamma)$   & $ \etap \to \gamma \pi^+ \pi^-$ & $ \etap \to \pi^+ \pi^- \eta(\gamma \gamma)$ \\
\hline
MDC tracking *  & 1.6 & 2.4 & 1.6 & 2.4  \\
PID * & 1.2 & 1.2 &  1.2 & 1.2 \\
Photon detection * & 0.6 & 1.2 &  0.6 & 1.2  \\
4C kinematic fit  & 0.5 & 0.5 & 0.5 & 0.5  \\
Veto of $\gamma$ conversion * & 1.0 & 1.0 & -- & -- \\
$\eta$ reconstruction & -- & 1.0 & -- & 1.0 \\
Form factor  & 0.4 & 0.2 & -- & --\\
Signal shape & 0.4 &  0.2  & -- & -- \\
Fit range and background shape  & 0.6 & 1.3 & -- & -- \\
Fixed peaking background  & 0.3 & 0.3 & -- & --\\
Number of $J/\psi$ events * & 0.5 & 0.5 &  0.5 & 0.5 \\
$\etap$ BFs  & 1.7 & 1.7 &1.7 & 1.7 \\
$\mathcal{B}(\ub \to e^+e^-)$*& -- & -- & --\footnotemark[1]/(0-14)\footnotemark[2] & --\footnotemark[1]/(0-14)\footnotemark[2] \\
Total & 3.1  & 4.0  & 2.8\footnotemark[1]/(2.8-14.3)\footnotemark[2] & 3.6\footnotemark[1]/(3.6-14.5)\footnotemark[2]\\
\hline
\hline
\end{tabular}
\end{center}
\footnotetext[1]{Uncertainties associated with the upper limit on $\mathcal{B}(\jpsi
\to \etap \ub)\times \mathcal{B}(\ub \to e^+e^-)$. }
\footnotetext[2]{Uncertainties associated with the upper limit on $\mathcal{B}(\jpsi
\to \etap \ub)$.}
\label{t_sys}
\end{table*}

\section{\boldmath Dark photon search result}
\begin{figure}[!htb]
\centering
\includegraphics[width=0.49\textwidth]{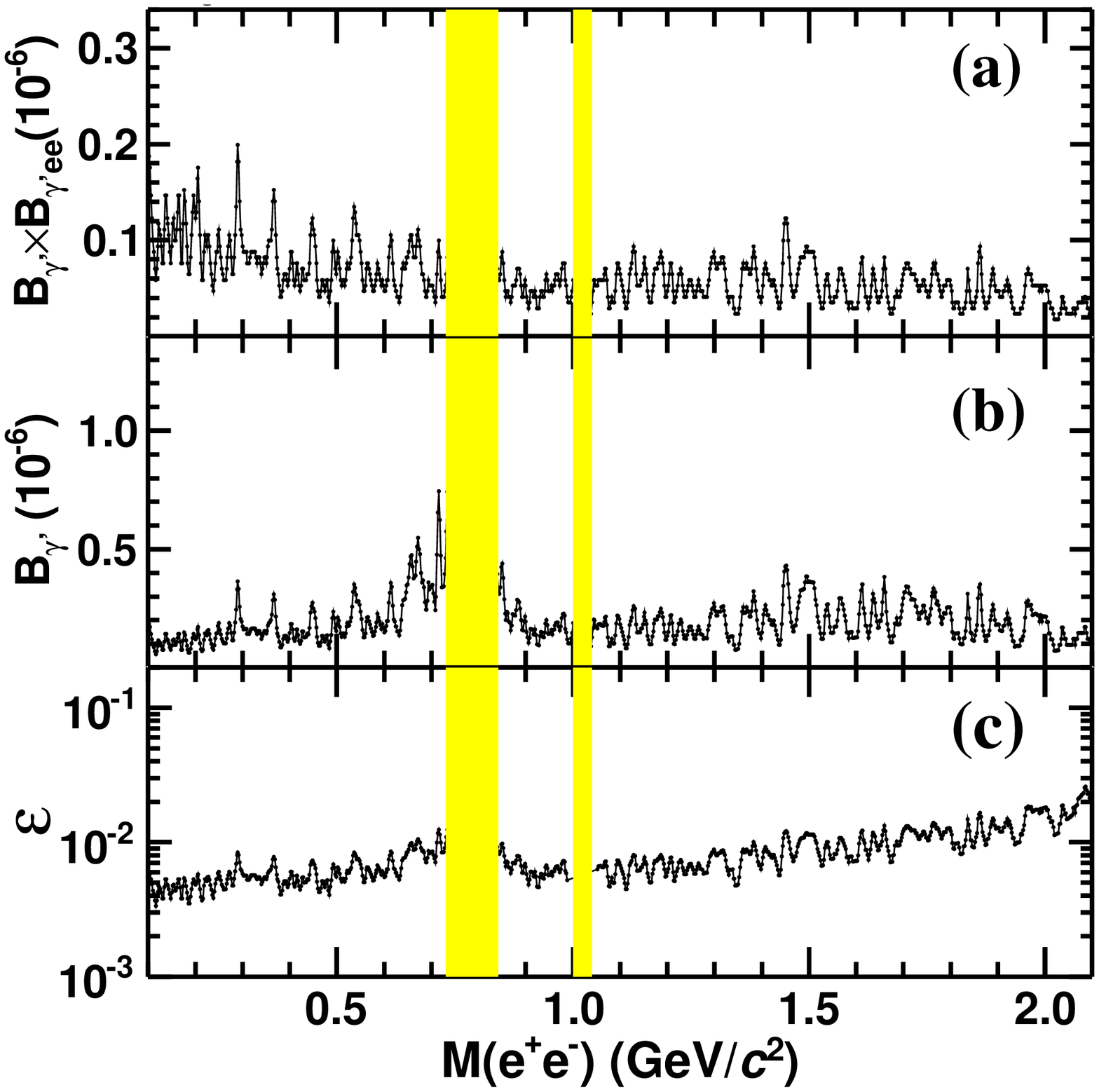}
\caption{ Upper limits at the 90\% C.L. for the BFs (a)
  $\mathcal{B}_{\ub}\times \mathcal{B}_{\rm \ub ee}$ and (b)
  $\mathcal{B}_{\ub}$, where $\mathcal{B}_{\ub}$ and $\mathcal{B}_{\rm
    \ub ee}$ are $\mathcal{B}(\jpsi \to \etap \ub)$ and
  $\mathcal{B}(\ub \to e^+e^-)$, respectively. (c) The exclusion limit
  at the 90\% C.L. on the kinematic mixing strength $\varepsilon$.}
\label{f_brul}
\end{figure}

We compute the upper limit on the BFs
$\mathcal{B}(\jpsi \to \etap \ub)\times \mathcal{B}(\ub \to e^+e^-)$
and $\mathcal{B}(\jpsi \to \etap \ub)$ at the 90\% C.L. using a
Bayesian method~\cite{r_pdg16}. The expected number of signal events
observed in the $i$th mode is calculated with
$N^i_{ sig} = N_{\jpsi} \cdot \mathcal{B}^{i}_{\etap \to {\rm F}}
\cdot \mathcal{B}(\jpsi \to \ub \etap) \cdot \mathcal{B}(\ub \to
e^+e^-) \cdot\mathscr{E}^i \cdot \delta^2$, where $N_{\jpsi}$ is the
number of $\jpsi$ events, $\mathcal{B}^{i}_{\etap \to {\rm F}}$ is the
BF of $\etap$ decay to final state F,
$\mathcal{B}(\jpsi \to \ub \etap)$ and $\mathcal{B}(\ub \to e^+e^-)$
are the BFs
%$\ub \to e^+e^-$ quoted from Ref.~\cite{Batell2009}, $\mathscr{E}^i$ is the detection efficiency determined from the
signal MC simulation, and $\delta=1.012$ is the electron tracking
efficiency correction factor. The likelihood value ($\mathcal{L}$), as
a function of product BF
$\mathcal{B}(\jpsi \to \etap \ub)\times \mathcal{B}(\ub \to e^+e^-)$,
is calculated as a product of $\mathcal{L}$ from mode I and mode II
with the method described in Ref.~\cite{r_combine}. The systematic
uncertainties, which have been discussed in Sec.~\ref{sec_syst},
are separately incorporated into the likelihood distribution as
correlated and uncorrelated terms. The upper limit
$\mathcal{B}^{\rm UP}$ on the product BF at the 90\% C.L. is
determined from the integral
$\int^{\mathcal{B}^{\rm UP}}_{0} \mathcal{L} d\mathcal{B}/
\int^{\infty}_0 \mathcal{L} d\mathcal{B} = 90\%$. The values of
$\mathcal{B}^{\rm UP}$ are plotted as a function of $m_{\ub}$ in
Fig.~\ref{f_brul}(a).  We also obtain the likelihood value as a
function of $\mathcal{B}(\jpsi \to \etap \ub)$ by taking into account
the BF $\mathcal{B}(\ub \to e^+e^-)$ and its corresponding
uncertainty~\cite{Batell2009}, and we compute the upper limit on the
$\mathcal{B}(\jpsi \to \etap \ub)$ at the 90\% C.L., as shown in
Fig.~\ref{f_brul}~(b). The upper limit at the 90\% C.L. on the BF
$\mathcal{B}(\jpsi \to \etap \ub)\times \mathcal{B}(\ub \to e^+e^-)$
ranges from $1.8\times10^{-8}$ to $2.0\times10^{-7}$ and that on
$\mathcal{B}(\jpsi \to \etap \ub)$ ranges from $5.7\times10^{-8}$ to
$7.4\times10^{-7}$.

The mixing strength $\varepsilon$ coupling $\ub$ and SM photon is
determined from the ratio of the BF
$\mathcal{B}(\jpsi \to \etap \ub)$ and that of the radiative process
$\mathcal{B}(\jpsi \to \etap \gamma)$ as~\cite{wang_gev}

$$
\frac{\mathcal{B}(\jpsi \to \etap \ub)}{\mathcal{B}(\jpsi \to \etap\gamma)} = \varepsilon ^2 |F(m^2_{\ub})|^2\frac{\lambda^{3/2} \left(m^2_{J/\psi},m^2_{\etap},m^2_{\ub}\right)}{\lambda^{3/2} \left(m^2_{J/\psi},m^2_{\etap},0\right)},
$$
where
$\lambda(m_1^2,m_2^2,m_3^2)=(1+\frac{m_3^2}{m_1^2-m_2^2})^2-\frac{4m_1^2m_3^2}{(m_1^2-m_2^2)^2}$;
$m_i$ is mass of a specific particle $i$, and $|F(m^2_{\ub})|^2 $ is
the $\jpsi$ to $\etap$ transition form factor as described in
Sec.~\ref{bes3nmc}, evaluated at $q^2$=$m_{\ub}^2$. The BF
$\mathcal{B}(\jpsi \to \etap \gamma)$ is taken from the
PDG~\cite{r_pdg16} value. The corresponding exclusion limit on the
mixing strength $\varepsilon$, which is shown in
Fig.~\ref{f_brul}~(c), ranges from $3.4 \times 10^{-3}$ to
$2.6 \times 10^{-2}$ depending on $m_{\ub}$.

\section{\boldmath Summary}

With a data sample of $(1310.6 \pm 7.0) \times 10^6$ $\jpsi$ events
collected by the BESIII detector, we measure the BF of the EM Dalitz
decay $\jpsi \to \etap e^+e^-$ with two dominant $\etap$ decay
modes. The combined result of $\mathcal{B}(\jpsi \to \etap e^+e^-)$ is
determined to be $(6.59\pm0.07\pm0.17) \times 10^{-5}$. This result is
compatible with the previous BESIII measurement~\cite{r_chuxkpee} and
the precision is greatly improved from 6\% to 3\%.

We also search for a dark photon via the decay chain
$\jpsi \to \etap \ub, \ub \to e^+e^-$ with the same two $\etap$ decay
modes. No significant signal of $\ub$ is observed, and we set upper
limits for the product BF
$\mathcal{B}(\jpsi \to \etap \ub)\times\mathcal{B}(\ub \to e^+e^-)$
and the BF $\mathcal{B}(\jpsi \to \etap \ub)$ at the 90\% C.L., which
range from $1.8\times10^{-8}$ to $2.0\times10^{-7}$ and
$5.7\times10^{-8}$ to $7.4\times10^{-7}$, respectively. The exclusion
limit on the mixing strength $\varepsilon $ between the SM photon and
dark photon varies in a range from $3.4 \times 10^{-3}$ to
$2.6 \times 10^{-2}$ depending on $m_{\ub}$. This is among the first
searches for the  dark photon in the charmonium decays.

\section*{\boldmath Acknowledgments}
The BESIII collaboration thanks the staff of BEPCII and the IHEP computing center for their strong support. This work is supported in part by National Key Basic Research Program of China under Contract No. 2015CB856700; National Natural Science Foundation of China (NSFC) under Contracts Nos. 11235011, 11335008, 11425524, 11625523, 11635010; the Chinese Academy of Sciences (CAS) Large-Scale Scientific Facility Program; the CAS Center for Excellence in Particle Physics (CCEPP); Joint Large-Scale Scientific Facility Funds of the NSFC and CAS under Contracts Nos. U1232105, U1332201, U1532257, U1532258; CAS Key Research Program of Frontier Sciences under Contracts Nos. QYZDJ-SSW-SLH003, QYZDJ-SSW-SLH040; 100 Talents Program of CAS; National 1000 Talents Program of China; INPAC and Shanghai Key Laboratory for Particle Physics and Cosmology; German Research Foundation DFG under Contracts Nos. Collaborative Research Center CRC 1044, FOR 2359; Istituto Nazionale di Fisica Nucleare, Italy; Koninklijke Nederlandse Akademie van Wetenschappen (KNAW) under Contract No. 530-4CDP03; Ministry of Development of Turkey under Contract No. DPT2006K-120470; National Science and Technology fund; The Swedish Research Council; U. S. Department of Energy under Contracts Nos. DE-FG02-05ER41374, DE-SC-0010118, DE-SC-0010504, DE-SC-0012069; University of Groningen (RuG) and the Helmholtzzentrum fuer Schwerionenforschung GmbH (GSI), Darmstadt; WCU Program of National Research Foundation of Korea under Contract No. R32-2008-000-10155-0

\end{document}